\documentclass[11pt]{article}
\usepackage[a4paper,margin=2.5cm]{geometry}
\usepackage{graphicx}
\graphicspath{{figures/}}
\usepackage{amsmath,amssymb}
\usepackage[numbers,sort&compress]{natbib}
\usepackage{booktabs}
\usepackage{tabularx}
\usepackage{ragged2e} 
\usepackage{float}
\usepackage{placeins}
\usepackage{xcolor}
\usepackage{enumitem}
\usepackage[most]{tcolorbox}
\usepackage[colorlinks=true,linkcolor=blue,citecolor=red,urlcolor=blue]{hyperref}
\usepackage{setspace}
\usepackage[font={small,stretch=1.0},labelfont=bf,skip=5pt]{caption}

\begin{document}

\title{Tensor Networks for Materials Simulation and Quantum Chemistry}

\author{Saeed S.~Jahromi$^{1,2}$\thanks{Corresponding author. Email: saeed.jahromi@multiversecomputing.com}%
, Augustine Kshetrimayum$^{1}$,\\
Roman Rausch$^{1}$, Sukhbinder Singh$^{3}$, Rom\'an Or\'us$^{1,2,4}$
\\[8pt]
{\small $^1$Multiverse Computing, San Sebasti\'an, Spain}\\
{\small $^2$Donostia International Physics Center, San Sebasti\'an, Spain}\\
{\small $^3$Multiverse Computing, Toronto, Ontario, Canada}\\
{\small $^4$Ikerbasque Foundation for Science, Bilbao, Spain}
}
\date{}

\maketitle

\begin{abstract}
Predicting the properties of correlated materials and molecules often requires a many-body description that goes beyond single-particle and mean-field approximations. Tensor networks address this challenge by exploiting the entanglement structure of quantum states to construct compact, systematically improvable representations. Although developed largely for idealized model Hamiltonians, these methods are increasingly being integrated into material-specific calculations. This Perspective introduces their underlying principles and examines how network geometry, bond-dimension truncation, and approximate contraction determine the accuracy and computational reach of these approaches. We survey applications across regimes of increasing complexity, ranging from ground-state properties of quantum magnets and itinerant-fermion systems to excitation spectra, nonequilibrium dynamics, finite-temperature and open systems, and molecular electronic structure. We then examine four routes connecting these methods to real materials, including experimentally constrained models, downfolded Hamiltonians, dynamical mean-field theory with tensor network impurity solvers, and quantum embedding. Throughout, we evaluate the maturity of existing capabilities and their principal limitations, synthesizing this assessment into a practical method-selection guide. We argue that reliable materials predictions require quantifying uncertainties introduced at each stage, from constructing material-specific Hamiltonians to extracting experimentally accessible observables. Finally, we identify the algorithmic advances and reproducible computational practices needed to establish tensor networks as a routine tool for predictive materials science and quantum chemistry.
\end{abstract}

\section{Introduction}\label{sec:intro}

Many of the most interesting questions in materials science are decided by small energy differences between very different states of matter. Whether a layered magnet orders or remains a spin liquid, whether a doped oxide superconducts or forms charge stripes, whether the transition-metal centre of a catalyst prefers a high-spin or a low-spin configuration, and how a moir\'e flat band responds when it is filled with electrons: in these settings competing many-body configurations lie close in energy, and the material chooses among them through correlations that no single-particle picture captures.

The workhorse of computational materials science, density functional theory (DFT), is remarkably successful at predicting structures, bands, phonons, and formation energies, and its extensions with Hubbard corrections or hybrid functionals stretch its reach into moderately correlated systems. Where configurations compete, however, the electronic structure must be treated as a genuine many-body problem, and each established many-body method has limitations in parts of this regime. Exact diagonalization, known in quantum chemistry as full configuration interaction, treats a finite Hilbert space without approximation but is limited by its exponential growth to a few dozen spins or orbitals, whether on a lattice or in a molecular active space \cite{Weisse2008}. Quantum Monte Carlo can give numerically exact results where a non-negative sampling weight exists, but many doped fermion and frustrated spin models suffer a sign problem whose severity can grow exponentially with size and inverse temperature \cite{Troyer2005}. Conventional single-reference coupled-cluster theory, a standard of quantum chemistry, is most reliable when one reference configuration dominates \cite{Bartlett2007}, and complete-active-space methods treat multireference problems exactly within a selected space whose size is bounded by the same exponential wall \cite{Olsen2011}. Dynamical mean-field theory (DMFT) captures local correlations in real materials while treating spatial correlations approximately \cite{Georges1996,Kotliar2006}. 

Tensor networks approach the problem from a different direction. Instead of asking how to sample or expand the full many-body state, they ask how much of the exponentially large Hilbert space a physical state actually occupies. Entanglement area laws explain why many physically relevant states admit compact descriptions: in important classes of local Hamiltonians, ground-state entanglement scales with the boundaries between regions. Tensor networks exploit this structure through small tensors connected by indices whose size, the bond dimension, controls the representational capacity \cite{Eisert2010,Hastings2007,Verstraete2006}. The density matrix renormalization group (DMRG), formulated by White in 1992, was the first algorithm to exploit this structure \cite{White1992,Schollwock2005}, and it was later understood as a variational method over matrix product states (MPS) \cite{Ostlund1995,Schollwock2011}. Since then the idea has grown into a family of representations for chains, trees, two-dimensional lattices, scale hierarchies, operators, thermal states, open-system dynamics, and partition functions \cite{Orus2014,Orus2019,Cirac2021,Banuls2023}.

Several properties make this family attractive for materials problems, and the established vocabulary places it precisely. Tensor network methods are \emph{nonperturbative}: they make no expansion in a small coupling and apply at the intermediate interaction strengths where materials are most interesting. They are \emph{controlled}: no term of the Hamiltonian is neglected, and the error of a calculation can be estimated from its convergence in the bond dimension and in the method-specific parameters of Box~1. They are \emph{systematically improvable}: as $\chi\to\infty$ the ansatz becomes exact, and wavefunction ansatzes give variational upper bounds when expectation values are evaluated exactly (although approximate contraction can compromise that guarantee). Beyond these three properties, tensor network wavefunction methods avoid the conventional sign problem, giving access to frustrated magnets and doped fermions, although their entanglement and optimization can still make them costly. They are agnostic to the type of degree of freedom and treat spins, fermions, bosons, phonon modes, and molecular orbitals within the same framework. They give direct access to the represented state, from which reduced density matrices, correlation functions, order parameters, and entanglement measures can be evaluated. The same representations serve ground states, excited states, real-time dynamics, finite temperature, and dissipation. The cost depends strongly on entanglement as well as system size, local degrees of freedom, and contraction geometry, so states with volume-law entanglement, long-time dynamics, and higher-dimensional Fermi surfaces remain hard.

For most of their history, tensor network methods were developed and refined on idealized model Hamiltonians, the Heisenberg and Hubbard models chief among them, and their development ran largely in parallel to the electronic-structure workflows of materials science and chemistry. That separation is now closing. DMRG is a standard tool for large active spaces in quantum chemistry \cite{Chan2011,Baiardi2020}, tensor network impurity solvers run inside dynamical mean-field theory for real compounds \cite{Bauernfeind2017,Linden2020}, downfolded models of nickelate superconductors and continuum models of moir\'e materials are solved on cylinders \cite{Qu2024,Soejima2020,Dong2023}, and ab initio embedding calculations reach the cuprate family \cite{Cui2022,Cui2025}. We examine that crossing, focusing on how the methods work, what they have achieved, where they fall short, and what is needed for them to become dependable components of predictive materials science. We argue that reliable material predictions depend on quantifying the uncertainty introduced at each stage, from the construction of the material-specific Hamiltonian to the extraction of experimentally accessible observables.

This Perspective is aimed at researchers who already use electronic-structure or quantum-chemistry methods and want to know what tensor networks can add to that toolbox. Section~\ref{sec:paradigm} introduces the ideas with a minimum of formalism and summarizes the sources of error a user must control (Figs.~\ref{fig:concepts} and \ref{fig:operations}, Box~1). Section~\ref{sec:regimes} follows the methods through regimes of increasing complexity, from spin chains to molecules, with landmark applications to real materials at each step (Fig.~\ref{fig:map}). Section~\ref{sec:real-materials} examines the four routes by which a tensor network calculation is connected to a specific compound (Fig.~\ref{fig:routes}) and how uncertainty accumulates along the chain from the material input to the measured observable. Section~\ref{sec:frontiers} condenses the state of the field into a method-selection table (Table~\ref{tab:master}), identifies the frontiers, and briefly places tensor networks in the context of quantum computing and machine learning. Throughout we label capabilities as \emph{established}, \emph{advanced specialist}, \emph{emerging}, or \emph{open}, so that a reader can judge what is routine and what still needs an expert. The examples that support these labels are illustrative rather than exhaustive: each was chosen to show what a capability delivers for a real material, and the literature behind every regime is far larger than a Perspective can survey.

\section{Tensor Network Fundamentals}\label{sec:paradigm}

\subsection{The exponential wall and the area law}\label{sec:arealaw}

\begin{figure}[!tb]
\centering
\includegraphics[width=\textwidth]{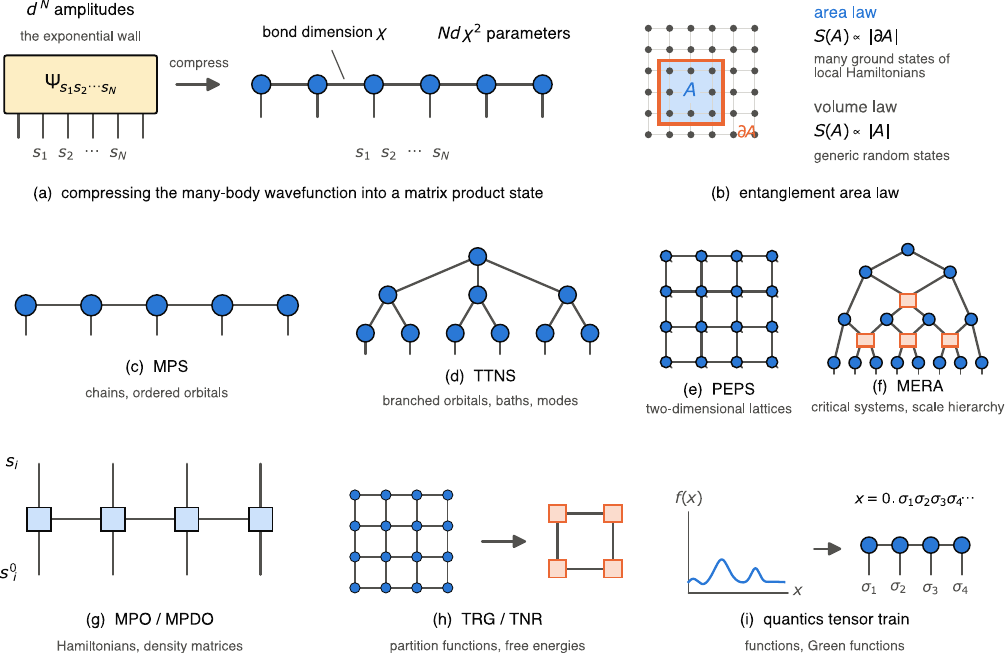}
\caption{Overview of tensor networks. (a) The $d^N$ amplitudes of a many-body wavefunction are replaced by a chain of $N$ small tensors connected by bond indices of dimension $\chi$; the bond dimension controls both the retained entanglement and the cost. (b) An important basis for compression is the area law obeyed by many ground states of local Hamiltonians: the entanglement entropy of a region $A$ scales with its boundary $\partial A$ rather than its volume. (c)--(f) The network geometry follows the entanglement structure: matrix product states (MPS) for chains and ordered orbitals, tree tensor network states (TTNS) for branched correlations, projected entangled-pair states (PEPS) for two-dimensional lattices, and the multiscale entanglement renormalization ansatz (MERA) for critical systems. (g) The same tensors represent operators and density matrices (MPO, MPDO), (h) partition functions that are coarse-grained by the tensor renormalization group (TRG, TNR). (i) Smooth functions such as Green functions can be discretized accurately as quantics tensor trains, matrix product states in which a continuous variable $x$ is encoded in its binary digits $\sigma_1\ldots\sigma_4$.}
\label{fig:concepts}
\end{figure}

A quantum state of $N$ spin-$1/2$ sites is specified by $2^N$ complex amplitudes, one for each configuration $s_1 s_2\ldots s_N$. For three hundred sites this exceeds the number of atoms in the observable universe, and a molecular active space of a hundred orbitals faces the same wall. Yet nature does not explore this space uniformly. A generic random pure state has entanglement entropy proportional to the volume of the smaller subsystem across a bipartition. Many ground states, and often the low-lying excited states, of local Hamiltonians behave differently: their entanglement entropy scales with the boundary between a region and its complement, which is the area law \cite{Eisert2010}. For gapped local one-dimensional systems, the entanglement across a cut is bounded independently of chain length, and ground states can be approximated efficiently by matrix product states \cite{Hastings2007,Verstraete2006}. Critical systems and states with a Fermi surface can violate the area law. Where compression works, a region needs to retain only a limited number of relevant states of its surroundings. Tensor networks are the representations built on that observation (Fig.~\ref{fig:concepts}).

\subsection{Matrix product states and the bond dimension}\label{sec:mps}

The simplest tensor network replaces the coefficient array by a product of matrices, one for each site,
\begin{equation}
|\Psi\rangle=\sum_{s_1\ldots s_N} A_1^{s_1}A_2^{s_2}\cdots A_N^{s_N}\,|s_1\ldots s_N\rangle.
\label{eq:mps}
\end{equation}
For open boundaries, $A_i^{s_i}$ has dimensions $\chi_{i-1}\times\chi_i$, with $\chi_0=\chi_N=1$ and $\chi_i\leq\chi$, so the product is a scalar; the physical index $s_i$ labels the local spin, orbital occupation, or vibrational state, and the matrix indices, called virtual or bond indices, are summed where neighbouring tensors meet. The bond dimension $\chi$ is the single most important parameter. For $\chi=1$ Eq.~\eqref{eq:mps} is a product state, the mean-field limit, and as $\chi$ grows the ansatz becomes exact. Its meaning is made precise by the Schmidt decomposition. Cutting the chain between sites $i$ and $i+1$ one can always write $|\Psi\rangle=\sum_{\alpha}\lambda_\alpha|L_\alpha\rangle|R_\alpha\rangle$ with orthonormal states on the left and on the right and non-negative Schmidt coefficients $\lambda_\alpha$. The number of non-zero coefficients is the bond dimension the exact state requires at that cut, and the entanglement entropy $S=-\sum_\alpha\lambda_\alpha^2\ln\lambda_\alpha^2$ is bounded by $\ln\chi$. Across this cut, retaining the $\chi$ largest coefficients gives the optimal rank-$\chi$ approximation in the state norm, and the discarded weight $\varepsilon=\sum_{\alpha>\chi}\lambda_\alpha^2$ measures the squared norm lost in that truncation, rather than a universal error bar for observables (Fig.~\ref{fig:operations}b). The number of parameters is $Nd\chi^2$ rather than $d^N$, and the algorithms scale as a low polynomial in $\chi$. Rapidly decaying Schmidt coefficients enable effective compression. Critical states, states with a Fermi surface, and states generated by long real-time evolution require a bond dimension that grows with system size or time and eventually defeat the compression \cite{Cirac2021}. In one dimension this limit is mild in practice: at a critical point the entropy grows only logarithmically with length, so a Heisenberg chain of hundreds of sites is solved to very high accuracy with modest $\chi$, and the residual finite-entanglement corrections obey scaling laws that can be extrapolated \cite{Tagliacozzo2008,Pollmann2009}.

\subsection{Geometry follows entanglement}\label{sec:geometry}

The chain of Eq.~\eqref{eq:mps} is one geometry among many, and the guiding rule is that the network should follow the structure of the entanglement (Fig.~\ref{fig:concepts}c--f). Matrix product states suit chains and any problem whose degrees of freedom can be ordered along a line: ladders, strips, nanoribbons, nanotubes, and the cylinders of Sec.~\ref{sec:chains}. Finite molecules are a less obvious but important case. Their orbital entanglement is bounded, and the canonical form of an MPS makes its optimization robust, so an MPS over molecular orbitals performs well in practice until the active space grows large enough that orbital correlations no longer fit a line (Sec.~\ref{sec:molecules}). Tree tensor network states replace the chain by a loop-free branching structure, which suits problems whose correlations separate into groups, such as branched molecular orbitals, the bath channels of an impurity, or groups of vibrational modes \cite{Shi2006ttn,Murg2010ttns,Nakatani2013ttns}; like the chain, a tree can be contracted exactly. Projected entangled-pair states (PEPS) place a tensor on every site of a two-dimensional lattice and connect neighbours by bonds, so that the area law is built into the ansatz and the thermodynamic limit is reached directly with infinite PEPS \cite{Verstraete2004peps,Jordan2008ipeps}; the price is that contracting a network with loops is exponentially hard in general and must be approximated \cite{SchuchPEPShardPRL2007, HaferkampPEPSPRR2020}. Isometric tensor network states constrain PEPS to a canonical-like form that makes contraction cheap at some cost in variational freedom \cite{Zaletel2020}. The multiscale entanglement renormalization ansatz (MERA) organizes tensors in layers that coarse-grain the system scale by scale, which is the natural representation at a critical point \cite{Vidal2007mera}. For materials, MPS and PEPS remain the workhorses. Trees have so far found use mainly in quantum chemistry and in impurity problems with several bath channels (Secs.~\ref{sec:molecules} and \ref{sec:dmft}). MERA, although the natural representation of a critical point, has not yet been applied competitively to a materials problem; its footprint in Table~\ref{tab:master} is correspondingly small. 

Further constructions extend these families: trees laid out to respect the area law simulate two-dimensional systems with exact contraction \cite{Tagliacozzo2009}, branching MERA describes highly entangled states beyond the area law \cite{Evenbly2014branching}, continuous matrix product states treat quantum fields without a lattice \cite{Verstraete2010cmps}, and matrix product and projected entangled-pair states can be sampled by Monte Carlo instead of contracted, trading contraction cost for a statistical error \cite{Sandvik2007,Schuch2008mc}.

The same tensors can represent objects other than pure states (Fig.~\ref{fig:concepts}g--i). A matrix product operator (MPO) encodes a Hamiltonian, including the long-range Coulomb interaction of a molecule \cite{Chan2016mpo}, and matrix product density operators, purifications, and projected entangled-pair operators (PEPO) encode thermal or dissipative mixed states \cite{Verstraete2004mpdo,Feiguin2005}. Coarse-graining methods such as the tensor renormalization group (TRG) and tensor network renormalization (TNR) act on classical or quantum partition functions written as tensor networks \cite{Levin2007trg,Xie2012hotrg,Evenbly2015tnr}. Quantics tensor trains represent smooth functions of continuous variables, such as Green functions on fine frequency grids, by encoding the coordinate in binary digits and compressing the resulting tensor \cite{Oseledets2011,Khoromskij2011,Shinaoka2023}; combined with tensor cross interpolation, which builds the compressed form from few function evaluations \cite{Ritter2024,NunezFernandez2025}, they have been used to learn Feynman diagrams \cite{NunezFernandez2022} and to perform fully numerical Hartree--Fock calculations \cite{Haubenwallner2025}; their use inside dynamical mean-field theory is described in Sec.~\ref{sec:dmft}.

\subsection{What the algorithms do}\label{sec:algorithms}

\begin{figure}[!tb]
\centering
\includegraphics[width=0.94\textwidth]{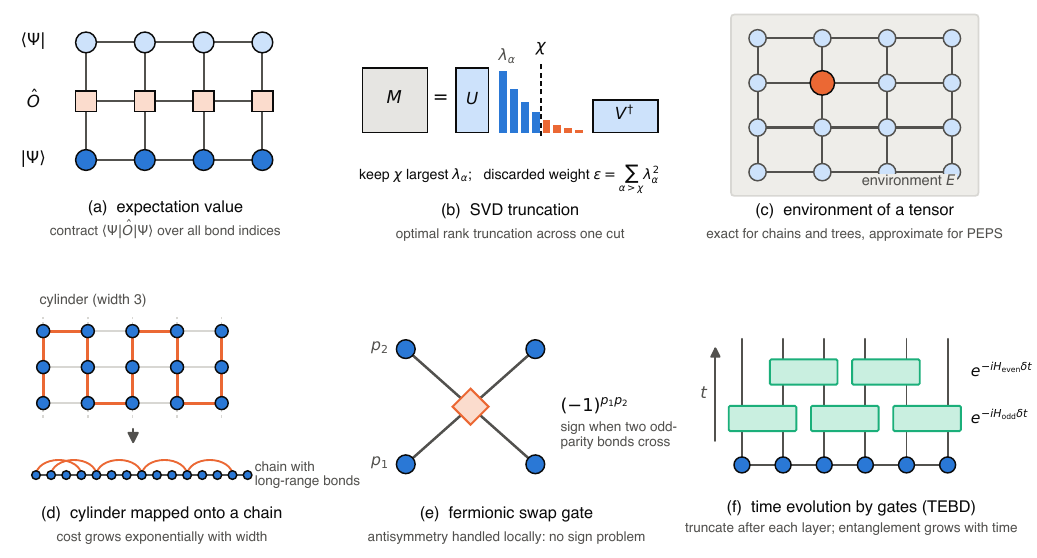}
\caption{The operations underlying every tensor network calculation. (a) An expectation value is the full contraction of the state, the operator, and the conjugate state. (b) Truncation by singular value decomposition keeps the $\chi$ largest Schmidt coefficients; the discarded weight $\varepsilon$ measures the norm lost at that cut and helps assess convergence. (c) To update or measure one tensor, the rest of the network is summarized in its environment, which is exact for chains and trees but must be approximated for PEPS. (d) A ladder or cylinder is simulated with one-dimensional methods by ordering the sites along a 1D snake path through the 2D lattice; local couplings become long-range bonds and the cost grows exponentially with the width. (e) Fermionic antisymmetry is handled locally by a swap gate that inserts a sign whenever two odd-parity bonds cross, so that antisymmetry costs only a polynomial overhead. (f) Time evolution applies layers of local gates and truncates after each layer; the entanglement, and hence the required bond dimension, grows with time.}
\label{fig:operations}
\end{figure}

Every tensor network algorithm either optimizes the tensors for a variational target or evolves them in real or imaginary time, and both rest on contraction, the evaluation of the network (Fig.~\ref{fig:operations}). Optimization finds the tensors that minimize the energy. DMRG does so by sweeping through the chain and solving an effective eigenvalue problem, of dimension set by $\chi$, at each site while the rest of the network is held fixed \cite{White1992,Schollwock2011}; the same sweep structure is used for trees, and gradient-based and automatically differentiated optimization has made variational PEPS practical \cite{Corboz2016,Liao2019,Naumann2025}. Evolution applies $e^{-iHt}$ or $e^{-\beta H}$ to a state, either by splitting the Hamiltonian into local gates and truncating after each layer, which is the time-evolving block decimation (TEBD) \cite{Vidal2004tebd}, or by projecting the exact evolution onto the tangent space of the ansatz, which is the time-dependent variational principle (TDVP) \cite{Haegeman2011,Haegeman2016,Paeckel2019}. Real time gives dynamics and spectra, imaginary time gives ground and thermal states; expanding the evolution operator or a spectral function in Chebyshev polynomials is an alternative that avoids a time step altogether (Sec.~\ref{sec:spectroscopy}) \cite{Holzner2011}. Contraction evaluates expectation values, overlaps, and partition functions by summing over all bond indices. For chains and trees this is exact and efficient. For PEPS the summation is approximated by representing the \emph{environment} of a site or a bond, that is, the contraction of everything else in the network, with a boundary matrix product state or a set of corner transfer matrices of finite dimension, the corner transfer matrix renormalization group (CTMRG) \cite{Nishino1996ctmrg,Orus2009ctm}, or, on arbitrary graphs, by belief propagation, which is cheap but neglects loop corrections \cite{Tindall2023bp}. How well that environment is captured is a second convergence parameter that a PEPS user controls alongside $\chi$; the cheap simple update uses a local approximation to it, whereas full and variational updates account for a more complete environment \cite{Jiang2008su,Corboz2016}. In all of these operations, symmetries can be built into the individual tensors, so that quantum numbers are conserved exactly and the cost is reduced, both for abelian groups and for non-abelian groups such as SU(2) \cite{McCulloch2002,Singh2010sym,Singh2011u1,Weichselbaum2012}.

The outputs of these operations are the quantities materials scientists need: energies and gaps, reduced density matrices and hence orbital occupations, correlation functions and order parameters, correlation lengths, entanglement spectra, dynamical correlation functions and the spectra derived from them, and thermodynamic functions.

\begin{tcolorbox}[colback=gray!5,colframe=gray!55,arc=2pt,boxrule=0.6pt,left=6pt,right=6pt,top=4pt,bottom=4pt,
title=Box 1. Controlling errors in a tensor network calculation,fonttitle=\bfseries\small,top=2pt,bottom=2pt]
\small
A converged bond dimension is necessary but not sufficient. The parameters that define a calculation, and that should be varied before a result is reported, fall into six groups.
\begin{itemize}[leftmargin=*,itemsep=0pt,topsep=1pt]
\item \textbf{Truncation.} Increase $\chi$ and check convergence of the target observables. For DMRG, energy extrapolation against discarded weight can be useful when a stable near-linear regime is observed; it is not a universal error estimate. Report the convergence sequence, not only the final value.
\item \textbf{Size and geometry.} Extrapolate in chain length, cylinder circumference, and aspect ratio, and compare open, periodic, and infinite boundary conditions. A small discarded weight cannot remove a geometry bias.
\item \textbf{Environment.} For PEPS, converge the boundary or corner-transfer-matrix dimension together with $\chi$, vary the unit cell, and compare simple, full, and variational updates.
\item \textbf{Ordering and topology.} For molecules and impurities, test orbital orderings, tree topologies, active spaces, and bath discretizations; use mutual-information maps to guide the choice.
\item \textbf{Time and frequency.} Converge the time step, maximum time, windowing, and broadening. Resolve the features of interest before comparison with the experimental resolution function.
\item \textbf{Sampling and optimization.} Report statistical errors where sampling is used, start from several initial states and symmetry sectors to expose metastability, and record the optimization history.
\end{itemize}
The general rule is that a small solver residual cannot compensate for an unsuitable network, an incomplete active space, or a wrong Hamiltonian.
\end{tcolorbox}

\subsection{Where tensor networks stand among other methods}\label{sec:positioning}

Tensor networks are one member of a toolbox, and their value is clearest when it is stated against the alternatives. Quantum Monte Carlo can be numerically exact, with statistical and algorithmic errors controlled, in sign-free settings such as many bosonic and unfrustrated spin models and the half-filled repulsive Hubbard model with hopping between opposite sublattices of a bipartite lattice, and it scales to large three-dimensional systems; tensor networks take over where the sign problem cannot be mitigated or resolved, in frustrated magnets and doped fermions, but at a cost that increases with dimensionality and entanglement. Neural quantum states share the variational philosophy and are not restricted by dimension, at the cost of stochastic optimization and the absence of a controlled truncation parameter; on standard two-dimensional benchmarks the two families are now closely matched, and shared metrics allow like-for-like comparison \cite{Carleo2017,Lange2024,Chen2024nqs,Wu2024vscore}. Conventional coupled-cluster theory is often more cost-effective for accurate calculations on single-reference molecules, and DMRG is the method of choice when the active space is large and multireference; the two can be combined. Dynamical mean-field theory addresses real materials directly, and tensor networks serve inside it as impurity solvers \cite{Bauernfeind2017,Wolf2015}. The competition between methods is therefore less important than their complementarity, and the strongest materials conclusions in this Perspective are those on which several methods agree.

\section{Tensor Networks Across Regimes of Correlated Matter}\label{sec:regimes}

\begin{figure}[!tb]
\centering
\includegraphics[width=\textwidth]{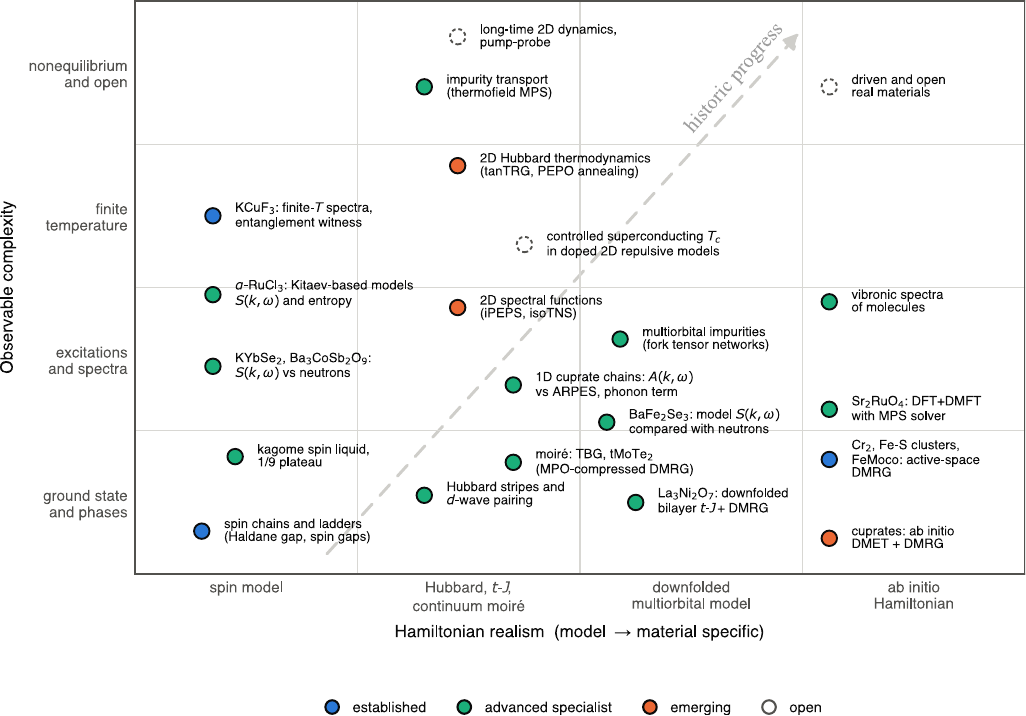}
\caption{A map of tensor network applications to materials. The horizontal axis orders the Hamiltonian from an idealized spin model to a material-specific ab initio description; the vertical axis orders the observable from a ground-state phase to nonequilibrium and open-system response. Each marker is a representative calculation discussed in Secs.~\ref{sec:regimes} and \ref{sec:real-materials}, coloured by the maturity of the capability it exemplifies. The historical development of the field runs along the diagonal from model ground states toward material-specific spectra and dynamics.}
\label{fig:map}
\end{figure}

We now follow the methods through problems of increasing complexity, from a spin chain to a molecule (Fig.~\ref{fig:map}). Each step raises the Hamiltonian realism, the observable complexity, or both, and each has its own natural network, its own landmark applications to real materials, and its own remaining difficulties. For each regime we state the question it addresses, the Hamiltonian or construction that encodes it, the tensor network and algorithm that suit it, what has been achieved, and how mature the capability is.

\subsection{Spin chains and ladders}\label{sec:chains}

The simplest correlated materials are those in which magnetic ions couple strongly along one crystallographic direction and weakly across it. Chains and ladders of copper, nickel, or cobalt ions, such as the spin-ladder cuprate Sr$_{14}$Cu$_{24}$O$_{41}$, realize the Heisenberg, XXZ, and spin-Peierls models, and their physics, the Haldane gap in integer-spin chains and the spin gaps in antiferromagnetic even-leg ladders, was among the first to be established quantitatively by DMRG, which also confirmed that higher half-integer-spin chains are gapless, as Haldane's semiclassical argument predicted but approximate methods had not settled \cite{White1992,White1993,White1994,Dagotto1996,Schollwock2005}. The relevant Hamiltonian is a spin model whose exchange couplings come from symmetry, from electronic-structure estimates, or from fits to susceptibility and neutron scattering data. Finite-system DMRG gives ground-state energies, gaps, correlation functions, and order parameters for chains of hundreds of sites; infinite-system variants work directly in the thermodynamic limit and remove boundary effects \cite{McCulloch2008,Zauner2018vumps,Vanderstraeten2019tangent}. Topological phases such as the Haldane phase are diagnosed from the degeneracy structure of the entanglement spectrum, a quantity readily available from a canonical MPS \cite{Pollmann2010, TuspinchainPRB2011, KshetrimayumspinchainPRB2015, Kshetrimayumspinchain2016}. For genuinely one-dimensional gapped systems these outputs are \emph{established}: rapid convergence with $\chi$ is often achievable and finite-size extrapolation is well developed. 

The contact with experiment is closest through spectroscopy, discussed in Sec.~\ref{sec:spectroscopy}, and a single example illustrates the level of detail that is now possible. For the Heisenberg-chain compound KCuF$_3$, finite-temperature DMRG described the inelastic neutron spectrum over a range of temperatures well enough that an entanglement witness, the quantum Fisher information, could be extracted from the measured data and compared with the calculation \cite{Scheie2021}; how such experimentally constrained models are built and fitted is the subject of Sec.~\ref{sec:models}.

Ladders and other quasi-one-dimensional lattices are handled by the same machinery once the sites are ordered along a path that snakes across the width (Fig.~\ref{fig:operations}d). Interactions that were local on the ladder become longer-range on the chain, and the bond dimension needed to describe entanglement across the width grows exponentially with the number of legs \cite{Stoudenmire2012}. This cylinder DMRG is the bridge to two dimensions. It is stable, variational, and mature, and it remains the reference method for two-dimensional models on cylinders of moderate circumference, at the cost of extrapolations in width and aspect ratio that must be reported alongside the bond dimension.

\subsection{Frustrated magnets in two and three dimensions}\label{sec:frustrated}

Frustration is the first step up in complexity. When competing exchange paths on a triangular or kagome lattice, or between first and second neighbours on a square lattice, cannot be satisfied simultaneously, classical order is suppressed and the ground state is selected by quantum fluctuations from a nearly degenerate manifold. Quantum Monte Carlo suffers a sign problem in exactly this situation, so tensor networks have become one of the main tools for frustrated magnetism in two dimensions. Two representations compete. Cylinder DMRG offers stable and well-understood variational optimization at circumferences up to roughly a dozen sites. Infinite PEPS work directly in the thermodynamic limit with native two-dimensional connectivity, using corner transfer matrices for the environment and simple, full, or variational updates for the optimization \cite{Jordan2008ipeps,Corboz2016,Naumann2024}; their observables depend on the environment dimension, the unit cell, and the optimization path, and these must be converged together with $\chi$. Tensor networks formulated on arbitrary lattices \cite{Jahromi2019universal}, including the ruby and triangle-honeycomb lattices \cite{Jahromi2018ruby}, fine-graining of lattices with large unit cells \cite{Schmoll2020finegrained}, and isometric networks \cite{Zaletel2020} extend the reach of the PEPS family. 

The kagome Heisenberg antiferromagnet shows both the power and the difficulty. Cylinder DMRG first found a spin-liquid ground state with a small spin gap consistent with $\mathbb{Z}_2$ topological order \cite{Yan2011}; later DMRG with different geometries found signatures of Dirac cones \cite{He2017kagome}, and infinite PEPS favoured a gapless spin liquid \cite{Liao2017}. The disagreement reflects distinct geometry and truncation effects, with cylinder width, boundary conditions, and bond dimension playing different roles, and the question is not fully settled. For frustrated systems, therefore, phase competition is an \emph{advanced specialist} capability, and agreement between methods is stronger evidence than any single record energy.

The comparison with experiment is nevertheless direct. Tensor network calculations characterized the nature of several magnetization plateaus of the kagome antiferromagnet \cite{Fang2023,PicotPRB2016,KshetrimayumkagomePRB2016}, and the $1/9$ plateau was subsequently observed in a kagome compound with an explicit connection to spin entanglement \cite{Jeon2024}. The double-layer kagome compound Ca$_{10}$Cr$_7$O$_{28}$ shows the route at the level of a full material model. Infinite projected entangled simplex states (iPESS), a generalization of iPEPS in which tensors sit on the simplices of the lattice \cite{XiangPESS2014}, reproduced the ground state seen by neutron scattering and its response to a magnetic field \cite{KshetrimayumAOP2020}, and a finite-temperature tensor network study of the same model resolved an earlier disagreement between the measured and calculated magnetization process \cite{KshetrimayummaterialPRB2024}. On the triangular lattice, cylinder DMRG combined with neutron scattering established the proximity of KYbSe$_2$ to a quantum spin liquid and identified signatures of fractionalization in its excitation continuum \cite{Scheie2024}. Matrix product state spectra showed that strong interactions can push a magnon out of a two-particle decay continuum instead of broadening it, which explains the anomalously sharp magnons observed in Ba$_3$CoSb$_2$O$_9$ \cite{Verresen2019}.

What makes these calculations credible is that the same Hamiltonian accounts for several observables at once, static susceptibility, ordered moment, dispersion, and spectral weight, over a range of temperatures and fields. When couplings are inferred from experiment, a parameter set that matches one dispersion but misses the temperature or polarization dependence is not yet a validated material model. Conclusions about spin-liquid or topological order likewise need diagnostics beyond the energy: correlation lengths, the scaling of the entanglement entropy, the topological entanglement entropy, or the response to flux insertion, each extrapolated in width or environment dimension.

Frustration is not confined to two dimensions, and the pyrochlore lattice, the three-dimensional archetype realized by many rare-earth and transition-metal oxides, has recently come within reach. DMRG with SU(2) symmetry on clusters of up to 128 sites characterized the ground state of the spin-$1/2$ pyrochlore Heisenberg antiferromagnet and found evidence for a spontaneous breaking of inversion symmetry \cite{Hagymasi2021}; the same machinery has followed its finite-temperature properties \cite{Schafer2020}, its magnetization process \cite{Hagymasi2022}, and the hard-hexagon crystals that populate its field-induced phases \cite{Schafer2023}. Three-dimensional PEPS, with an efficient contraction scheme for layered systems, extend the two-dimensional ansatz to the third dimension \cite{Vlaar2021,Vlaar2023}, and bilayer models have also been addressed \cite{Samimi2026}. Three-dimensional frustrated magnets are therefore an \emph{emerging} capability: cluster sizes are modest, the extrapolation to the bulk is delicate, and the accessible bond dimensions are far below those of chains, but these results show that the sign-free reach of tensor networks is not limited to planar geometries.

\subsection{Itinerant fermions and superconductivity}\label{sec:fermions}

Adding charge motion to spin brings us to the Hubbard and $t$--$J$ models, the minimal descriptions of doped Mott insulators, and to the physics of cuprate and nickelate superconductors and of moir\'e materials. Fermions introduce a technical challenge that tensor networks handle cleanly. In one dimension the Jordan--Wigner transformation maps fermions onto spins with string operators, and DMRG treats the resulting strings within the operator representation. In two dimensions a Jordan--Wigner ordering introduces nonlocal strings, while fermionic networks can track the signs locally: fermionic PEPS attach a parity to every bond and insert a sign wherever two odd-parity bonds cross, so that the antisymmetry is handled exactly with only a polynomial overhead (Fig.~\ref{fig:operations}e) \cite{Corboz2010fpeps,Kraus2010fpeps}. The entanglement cost remains, and doped fermion models often require greater bond dimensions than comparable spin models, so the accessible bond dimensions and widths are smaller.

The two-dimensional Hubbard model has become the arena in which these methods are stress-tested against one another. A community benchmark established reference energies and order parameters across the phase diagram by comparing DMRG, PEPS, diagrammatic and auxiliary-field Monte Carlo, and embedding methods \cite{LeBlanc2015}. For the underdoped region at one-eighth hole doping, the agreement of cylinder DMRG, infinite PEPS, constrained-path Monte Carlo, and density matrix embedding on a stripe ground state settled a long debate for that regime in the phase diagram \cite{Zheng2017}; the stripe is a genuine many-body prediction and not a mean-field artefact. With a next-nearest-neighbour hopping included, DMRG combined with auxiliary-field Monte Carlo on wide cylinders found $d$-wave superconductivity coexisting with partially filled stripes on the hole-doped side and weaker superconductivity with N\'eel correlations on the electron-doped side, in qualitative accord with the cuprates \cite{Xu2024}. Infinite PEPS showed that the uniform $d$-wave and stripe states of the $t$--$J$ model are energetically competitive \cite{Corboz2014}, DMRG on the $t$--$t'$--$J$ model mapped the competition between stripes and superconductivity \cite{Gong2021}, finite-temperature tensor networks followed how pairing evolves from charge clusters to stripes \cite{Sinha2026}, and finite fermionic PEPS have recently reported variational energies below those of the largest DMRG calculations on eight-leg open Hubbard ladders \cite{Liu2025fpeps}, evidence that the two-dimensional ansatz is becoming competitive on its own terms. A recent review takes stock of the whole computational programme \cite{Qin2022}.

Real materials enter through downfolded models. A bilayer $t$--$J$--$J_\perp$ model derived from the electronic structure of the pressurized nickelate La$_3$Ni$_2$O$_7$ has been solved with DMRG \cite{Qu2024}, and continuum models of moir\'e materials, compressed into matrix product operators \cite{Soejima2020}, have yielded a composite Fermi liquid at zero field in twisted MoTe$_2$ \cite{Dong2023}; a Wannier-projected lattice model of the same material has likewise yielded a map of fractional Chern insulators and quantum anomalous Hall crystals \cite{Chen2026}. How these models are constructed, and where their uncertainty sits, is the subject of Secs.~\ref{sec:models} and \ref{sec:downfolding}.

For itinerant fermions, ground-state phase competition is an \emph{advanced specialist} capability, and the identification of superconductivity is the hardest case. An energy minimum is not enough: algebraic pair-field correlations on a fixed-width cylinder indicate quasi-long-range order, while establishing two-dimensional superconducting order requires width scaling and supporting diagnostics; competing charge and magnetic orders must be tracked simultaneously, and the extrapolation in width, bond dimension, or environment must be reported for the correlations and not only for the energy. Controlled superconducting transition temperatures for the doped two-dimensional repulsive models discussed here remain an \emph{open} challenge, and in strictly two-dimensional systems the finite-temperature transition is of Berezinskii--Kosterlitz--Thouless type, so a transition temperature must be located from the onset of quasi-long-range order rather than from a true order parameter.

\subsection{Excitations, spectroscopy, and dynamics}\label{sec:spectroscopy}

A converged ground state is where most experimental comparisons begin rather than end. Inelastic neutron scattering, angle-resolved photoemission, optical absorption, resonant x-ray scattering, and pump-probe experiments all measure dynamical correlation functions, and tensor networks provide two routes to them. In the time domain the state is perturbed and evolved, the correlation function $C(t)=\langle\Psi|B(t)A|\Psi\rangle$ is recorded, and a Fourier transform gives the dynamical structure factor $S(k,\omega)$ or the single-particle spectral function $A(k,\omega)$ \cite{White2008spectral,Paeckel2019}. Modern integrators, the time-dependent variational principle with controlled bond expansion and fast QR-based block decimation, reduce the cost per time step and so extend the reachable time window within a fixed budget \cite{Li2024cbe,Unfried2023}. In the frequency domain, dynamical DMRG solves for the resolvent at each frequency \cite{Jeckelmann2002} while Chebyshev expansions deliver the whole spectrum from a single sequence of MPS operations at a resolution set by the expansion order \cite{Holzner2011}. For isolated quasiparticles, a variational excitation ansatz built on the ground-state MPS or PEPS gives dispersion relations directly, without any time evolution \cite{Haegeman2012exc,Vanderstraeten2015exc}.

For time-domain methods, a principal limiting factor is entanglement growth; frequency-domain and excitation-ansatz methods instead face resolution, variational-space, and contraction limitations. After a local perturbation the bond dimension needed to represent the state grows with time, so the accessible time window is finite and the frequency resolution is bounded by roughly its inverse. Finite-time spectra, and for Chebyshev expansions finite-order spectra, therefore carry numerical broadening, which must be controlled before convolution with the experimental resolution; linear prediction or finite-time extrapolation may extend the window but introduces an additional approximation. For one-dimensional models these calculations are \emph{established}, at zero temperature and, with the thermal constructions of Sec.~\ref{sec:thermal}, at finite temperature, and they have become a standard interpretive tool for neutron scattering on chain compounds (KCuF$_3$, Sec.~\ref{sec:chains}). On ladders and cylinders and in multiorbital impurity models they are an \emph{advanced specialist} capability, as the KYbSe$_2$ and Ba$_3$CoSb$_2$O$_9$ studies of Sec.~\ref{sec:frustrated} illustrate \cite{Scheie2024,Verresen2019}, and the finite-temperature spectrum of a single hole in a doped antiferromagnet has recently been resolved as a magnetic polaron \cite{Guthardt2025}. Two-dimensional spectra in the thermodynamic limit are \emph{emerging}: spectral functions have been computed directly with infinite PEPS \cite{AriasEspinoza2024}, isometric networks evolve two-dimensional spin systems in real time \cite{Lin2022}, and two- and three-dimensional tensor networks have followed the dynamics of disordered quantum systems to times that challenge quantum hardware \cite{Tindall2026}.

Nonequilibrium dynamics broadens the question from linear response to quenches, periodic driving, and pump-probe protocols, in which a Hamiltonian parameter is changed suddenly or a field is applied and the system is followed as it relaxes. The same time-evolution methods apply, and for one-dimensional models they have delivered thermalization dynamics, light-induced transient states, and transport coefficients. For realistic materials the difficulty is twofold: the entanglement grows without the bound that low energy provides, and the pulse, the dissipation channels, and the coupling to phonons must all enter the model. In two dimensions, infinite PEPS time evolution has followed many-body localized dynamics and disordered time crystals \cite{Kshetrimayum2020mbl,KshetrimayumTCPRB2021}. Long-time dynamics in two dimensions and predictive simulation of driven materials remain \emph{open}.

\subsection{Finite temperature and open systems}\label{sec:thermal}

Experiments are performed at finite temperature and often on systems coupled to an environment, and tensor networks represent both situations with mixed-state constructions. Purification doubles the system with an ancilla for every site so that a thermal state becomes a pure state on the enlarged space, obtained by imaginary-time evolution from infinite temperature \cite{Feiguin2005}. Minimally entangled typical thermal states (METTS) replace the ancilla by sampling over product states evolved in imaginary time, trading ancilla cost for statistical error \cite{White2009metts,Stoudenmire2010}. Matrix product density operators represent the density matrix itself as an operator network \cite{Verstraete2004mpdo}, at the risk of losing positivity under truncation. The exponential tensor renormalization group (XTRG) doubles the inverse temperature at each step by squaring the density operator, which reaches low temperature in few steps \cite{Chen2018xtrg}. For a one-dimensional local Hamiltonian, thermodynamic quantities, thermal correlations, and finite-temperature spectra are \emph{established} and have been decisive in the neutron-scattering comparisons above. Transport through quantum-impurity models at finite temperature and finite bias has been obtained with thermofield matrix product states \cite{Schwarz2018}.

In two dimensions the tools are younger. Projected entangled-pair operators annealed from infinite temperature give thermal states of two-dimensional quantum lattices \cite{Kshetrimayum2019thermal}, thermal isometric networks extend the isometric construction to density operators \cite{Kadow2023}, and a tangent-space thermal tensor network method (tanTRG) reached the doped two-dimensional Hubbard model at low temperatures where conventional determinant Monte Carlo is restricted by the sign problem \cite{Li2023tantrg}. Partition-function methods take a different route: the tensor renormalization group and its higher-order and tensor-network-renormalization refinements coarse-grain a classical or Euclidean quantum partition function written as a tensor network, yielding free energies, specific heats, and thermal phase boundaries without ever constructing a density matrix \cite{Levin2007trg,Xie2012hotrg,Evenbly2015tnr,Evenbly2017tnr}. An infinite-lattice thermal tensor network has also been applied to bosons in three-dimensional optical lattices \cite{Jahromi2020thermal}. Two-dimensional thermal phase behaviour is \emph{emerging}: the accessible temperatures, widths, and bond dimensions remain restricted, and the errors of the different constructions, bond truncation, sampling variance, positivity, and coarse-graining truncation, must be controlled separately.

Open quantum systems complete the picture. A Markovian environment enters through a Lindblad master equation, and the steady state of a two-dimensional dissipative lattice can be found by evolving a projected entangled-pair operator under the Liouvillian \cite{Kshetrimayum2017steady}. A non-Markovian environment with memory, the case relevant to vibronic dynamics and to electronic excitations coupled to phonons, is handled either by mapping the bath onto a chain of modes that is included in the matrix product state \cite{Chin2010chain,Prior2010,Schroder2019}, or by representing the influence functional itself as a tensor network in time, which is the idea behind the time-evolving matrix product operator (TEMPO) and process-tensor approaches \cite{Strathearn2018tempo,Jorgensen2019}. Steady states and non-Markovian dynamics of model open systems are an \emph{advanced specialist} capability; realistic open-device modelling for specific compounds remains \emph{open}, limited by the growth of entanglement and bath memory and by the uncertainty in the environmental model itself.

\subsection{Molecules and quantum chemistry}\label{sec:molecules}

Molecules and multiorbital clusters raise the complexity in a different direction. The degrees of freedom are molecular orbitals with no natural geometry, coupled by a Hamiltonian whose number of long-range Coulomb terms grows as the fourth power of the number of orbitals. The chemist's standard tool for strong correlation, the complete active space self-consistent field method, diagonalizes the Hamiltonian exactly within a chosen set of active orbitals, at an exponential cost that limits the active space to about eighteen to twenty orbitals. DMRG lifts this limit to fifty to a hundred orbitals by writing the active-space wavefunction as an MPS over orbitals \cite{Chan2011,Baiardi2020}. The ordering of the orbitals along the chain matters, because the bond dimension must carry the entanglement between all orbitals on either side of a cut; mutual-information measures between orbitals quantify that entanglement and guide both the ordering and the choice of active space \cite{Legeza2003,Rissler2006,Barcza2011}. Orbital optimization gives DMRG-SCF, the direct analogue of CASSCF \cite{Zgid2008,Ghosh2008}; spin adaptation reduces the cost and targets spin states exactly \cite{Sharma2012spin}; the ab initio Hamiltonian is handled efficiently as a matrix product operator \cite{Chan2016mpo}; and mature open-source codes such as Block2, CheMPS2, and QCMaquis make these calculations available to non-specialists \cite{Zhai2023block2,Wouters2014,Szenes2025}. Excited states are reached by state-averaged or targeted DMRG \cite{Dorando2007}. Model Hamiltonians on molecular graphs are a complementary use of the same machinery: the Hubbard model on the sixty-site C$_{60}$ cage, at the edge of what DMRG can converge, probes pair binding and Hund's-rule physics in fullerenes \cite{Rausch2026c60}.

The landmark applications are transition-metal systems whose multireference character defeats single-reference methods. DMRG resolved the electronic structure of the Mn$_4$CaO$_5$ cluster of photosystem II \cite{Kurashige2013}, computed the low-energy spectrum of iron--sulfur clusters directly from the many-electron Hamiltonian \cite{Sharma2014fes}, characterized the electronic complexity of the FeMo cofactor of nitrogenase \cite{Li2019femoco}, and, with a near-exact treatment of both static and dynamic correlation, closed the long controversy over the binding curve of the chromium dimer \cite{Larsson2022cr2}. Despite its success for molecular simulation, two limitations follow from the one-dimensional structure of the ansatz. First, DMRG captures correlation within the active space, including the static correlation that motivates its choice, while dynamic correlation involving the remaining orbitals must be added by perturbation theory, tailored coupled cluster, or related schemes \cite{Kurashige2011caspt2,Guo2016nevpt2,Veis2016tcc}. Second, the orbitals of a three-dimensional molecule are not a chain, and orbital correlations need not follow a one-dimensional pattern, so reaching chemical accuracy can require rapidly increasing bond dimensions as the active space grows; practical limits depend on the system and orbital basis.

Tree tensor networks address the second limitation by following branched correlation patterns rather than a line \cite{Murg2010ttns,Nakatani2013ttns}. The three-legged tree, T3NS, keeps the individual tensors small and has been applied to the competing bis($\mu$-oxo) and peroxo structures of the [Cu$_2$O$_2$]$^{2+}$ core, a classic multireference benchmark \cite{Gunst2018t3ns}. A tree lowers the bond dimension only when the orbital correlations are themselves tree-like, which is a question about the mutual-information graph and not about the molecule's shape, and the exact contraction of a tree costs more per unit of bond dimension than a chain, so the trade-off must be examined for each system. Fork tensor product states apply the same logic to impurity problems (Sec.~\ref{sec:dmft}) \cite{Bauernfeind2017}.

Magnetic molecules with many spin centres are a related class in which the Hamiltonian is a Heisenberg model on a molecular graph rather than an ab initio active space. DMRG reached the Keplerate Mo$_{72}$Fe$_{30}$ early on \cite{Exler2003}, has solved the Heisenberg model on the C$_{60}$ geometry \cite{Rausch2021c60}, a case that also serves as a benchmark against neural quantum states \cite{Szabo2024}, and, with SU(2) symmetry, has established the low-spin ground states of the Mn$_{70}$ and Mn$_{84}$ wheels, the largest single-molecule magnets synthesized to date \cite{Rausch2025mn84}.

Vibrational and vibronic problems are the natural extension. Replacing orbitals by vibrational modes gives anharmonic vibrational spectra, trees adapted to the coupling structure of the modes give converged spectra of polyatomic molecules such as acetonitrile \cite{Larsson2019ttns}, and combining electronic and vibrational degrees of freedom gives vibronic spectra, non-adiabatic dynamics, and models of electron--phonon coupling \cite{Baiardi2019dyn,Szenes2025}. For extended electron--phonon systems such as the Holstein and Peierls models, the bosonic Hilbert space of each phonon mode is compressed by an optimized local basis \cite{Zhang1998lbo}. Active-space energies and properties of molecules are \emph{established}; tree and vibronic calculations are \emph{advanced specialist}; predictive coupled electron--nuclear dynamics in extended solids is \emph{open}.

\section{From Models to Real Materials}\label{sec:real-materials}

\begin{figure}[!tb]
\centering
\includegraphics[width=\textwidth]{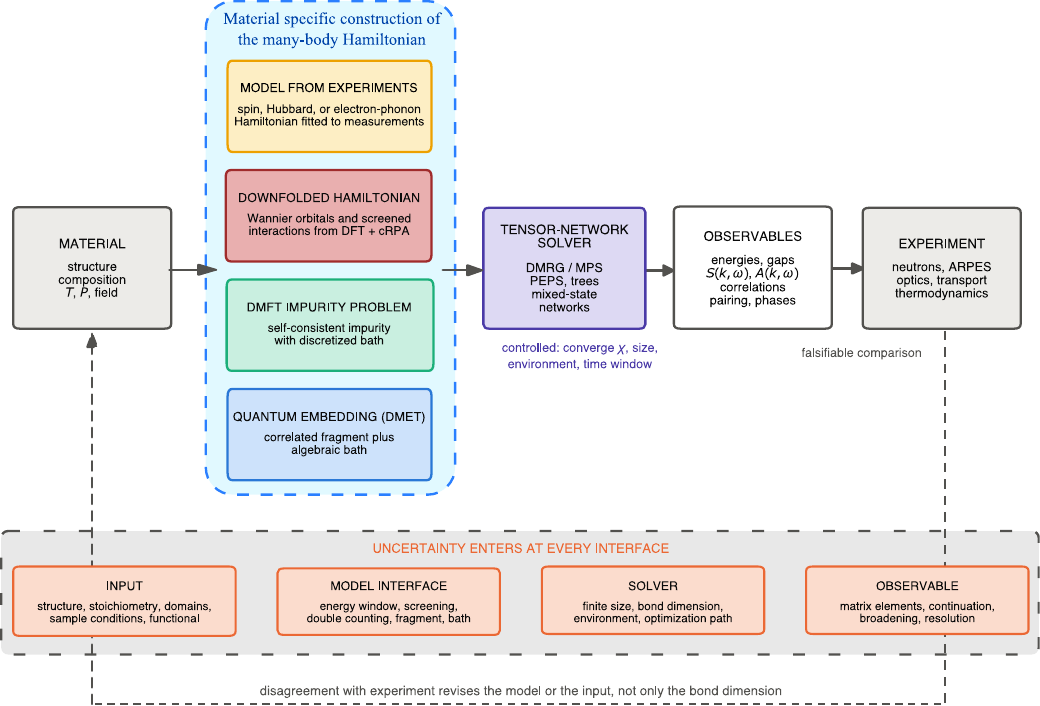}
\caption{Routes from a real material to an experimentally testable tensor network prediction. Structure, composition, and sample conditions feed one of four material-specific constructions: a model Hamiltonian constrained by experiment, a downfolded Hamiltonian from electronic structure, a self-consistent impurity problem from dynamical mean-field theory, or an embedded fragment with an algebraic bath. A tensor network solver returns the state or correlation function from which observables are extracted and compared with experiment. Uncertainty enters at every interface (orange), and disagreement with experiment should feed back to the model or the input rather than only to the solver.}
\label{fig:routes}
\end{figure}

Every calculation in the preceding section solved a Hamiltonian. Turning that solution into a statement about a specific compound requires five linked steps: specify the structure and the experimental conditions, construct a material-specific Hamiltonian or embedded problem, solve it, extract a measurable observable, and compare with experiment. Figure~\ref{fig:routes} shows the four routes through the second step. In each of them the tensor network is a controlled and systematically improvable solver for a chosen model, while the mapping from the real material to that model remains approximate and material specific.

The route is chosen by the question, not by the method. For the spectrum or thermodynamics of a magnetic insulator that neutron, susceptibility, or specific-heat data already constrain, the shortest path is an effective model whose form is fixed by symmetry and whose parameters are fitted to those data (Sec.~\ref{sec:models}). For a ground-state phase or a pairing tendency of an itinerant system with a few well-separated low-energy bands, the natural path is to downfold the electronic structure onto a lattice model (Sec.~\ref{sec:downfolding}). For one-particle spectra and other quantities dominated by local multiplet physics in a multiorbital $3d$, $4d$, $5d$, or $f$-electron compound, dynamical mean-field theory is the standard path, with a tensor network as one of its impurity solvers (Sec.~\ref{sec:dmft}). For energetics, local moments, and trends across a family of compounds whose correlated region can be localized in space, quantum embedding of a fragment in a periodic ab initio calculation stays closest to first principles (Sec.~\ref{sec:embedding}).

\subsection{Experimentally constrained models}\label{sec:models}

The shortest route begins with an effective spin, Hubbard, or electron--phonon Hamiltonian whose form is fixed by symmetry and chemistry and whose parameters come from measured dispersions, susceptibilities, or structural trends. For a magnetic insulator, the point-group symmetry of each exchange bond and the spin--orbit coupling on the magnetic ion decide which components of the exchange tensor are allowed. A Cu$^{2+}$ ion with quenched orbital moment gives a nearly isotropic Heisenberg exchange, so the chain and ladder cuprates of Sec.~\ref{sec:chains} need one or two parameters. A $4d$ or $5d$ ion with strong spin--orbit coupling on a honeycomb lattice, as in $\alpha$-RuCl$_3$ and the iridates, gives bond-dependent Kitaev, $\Gamma$, and $\Gamma'$ terms alongside the Heisenberg exchange \cite{Rau2014,Takagi2019}, and a rare-earth Kramers doublet on a triangular lattice needs four anisotropic constants \cite{Li2016aniso}. In these cases the crystal-field and spin--orbit level scheme that defines the effective spin-$1/2$ is itself an input, usually measured by neutron or optical spectroscopy. The model thus has one parameter for KCuF$_3$ and five or six for a spin--orbit-coupled honeycomb magnet: small in parameters, but not in entanglement.

The parameters are fixed in stages. The usual first pass is linear spin-wave theory fitted to the magnon dispersion from inelastic neutron scattering. This can fail in strongly quantum magnets and spin-liquid candidates whose spectrum is dominated by a continuum, motivating tensor networks as forward models for fitting: the dynamical structure factor is computed at the experimental temperature with the methods of Secs.~\ref{sec:spectroscopy} and \ref{sec:thermal}, the whole intensity map rather than the peak positions is compared with the data, and the parameters are adjusted until the map is reproduced. In KYbSe$_2$, Onsager reaction-field fits constrained the couplings, while cylinder tensor network spectra supported the model and the interpretation of proximity to a spin liquid \cite{Scheie2024}. For KCuF$_3$, finite-temperature DMRG provided the comparison for an entanglement witness extracted from the measured spectrum \cite{Scheie2021}. For $\alpha$-RuCl$_3$, MPS spectra of the Kitaev--Heisenberg model on cylinders \cite{Gohlke2017} and finite-temperature exponential tensor renormalization group calculations of a Kitaev--Heisenberg--$\Gamma$--$\Gamma'$ model that reproduce the double-peak specific heat \cite{Li2021RuCl3} play the same role for the neutron continuum and the two-step release of magnetic entropy, and together they show that the position of the material in its symmetry-allowed parameter space is still not uniquely determined.

Moir\'e materials are the itinerant version of the same route. Their natural description is a continuum model \cite{Bistritzer2011,Wu2019} whose single-particle parameters come from electronic structure but whose interaction scale is set by a dielectric constant that depends on the substrate, the gate distance, and the screening by remote bands, and is in practice fixed by experiment or left free. The Hamiltonian is projected onto the flat bands, compressed into an MPO, and solved by cylinder DMRG \cite{Soejima2020}. Here the characterization of the sample matters as much as the convergence of the solver: a heterostrain of one or two tenths of a percent, common in real devices, changes the ground state that DMRG selects in twisted bilayer graphene at charge neutrality \cite{Parker2021}, so the strain of the specific sample is a model parameter.

The electron--phonon case shows the route working in reverse, with the calculation revealing a term missing from the model. Angle-resolved photoemission on doped one-dimensional cuprate chains required a near-neighbour attraction comparable to the hopping, absent from the standard model of the cuprates \cite{Chen2021chain}, and variational non-Gaussian exact diagonalization \cite{Wang2021phonon} followed by time-dependent DMRG \cite{Tang2023} for a Hubbard--extended-Holstein chain then showed that extended electron--phonon coupling can produce an attraction of the required size. Finite-temperature DMRG now also delivers the thermodynamics, spectral functions, and optical conductivity of such electron--phonon models \cite{Jansen2020,Jansen2023}, bringing the comparison to the temperatures at which photoemission and optics are measured. A well-converged one-dimensional solution makes this a sensitive test of the Hamiltonian, provided uncertainties in sample conditions and in the conversion to the measured spectrum are also assessed. 

For magnetic insulators with a validated model, spectra and thermodynamics are \emph{established} in one dimension and \emph{advanced specialist} in two; ground-state phases of moir\'e and electron--phonon models are \emph{advanced specialist}; parameter inference with a tensor network as the forward model is \emph{emerging}. In every case credibility requires both a converged solver and a model validated against observables beyond those used to fit it.

\subsection{Downfolded Hamiltonians}\label{sec:downfolding}

Downfolding replaces the fit by a derivation. It starts from a crystal structure and a density functional calculation, selects a low-energy window of bands, constructs localized Wannier orbitals within it, and estimates the screened interactions between them \cite{Marzari2012,Aryasetiawan2004,Imada2010}. The structure comes from diffraction, the exchange--correlation functional sets the band widths, and maximally localized Wannier functions within the window give the hopping matrix $t_{ij}$ directly \cite{Marzari2012,Pizzi2020}. The interactions are estimated with the constrained random-phase approximation, which excludes polarization processes wholly within the selected low-energy subspace and which returns a frequency-dependent on-site $U$, intersite $V$, and Hund's $J$. The static limit of the resulting Hubbard-like model is then solved by DMRG or PEPS with high accuracy, and in the single-band repulsive Hubbard model at large $U/t$, a strong-coupling expansion gives superexchange $J=4t^2/U$ and a low-energy space without double occupancy. A further truncation gives the usual $t$--$J$ model, omitting terms such as three-site hopping away from half filling. Such reductions simplify the degrees of freedom but do not guarantee lower entanglement, and their validity depends on the relevant energy scales.

The choice of window is the decision with the largest consequences. For the cuprates, a single Cu $d_{x^2-y^2}$ band gives the one-band Hubbard model of Sec.~\ref{sec:fermions}, while retaining the oxygen $p_\sigma$ orbitals gives the three-band Emery model \cite{Emery1987} with an explicit charge-transfer energy, Cu--O hopping, and interactions on the retained orbitals. DMRG on ladders of the three-band model reproduces the asymmetry between hole and electron doping, the stripe formation, and the distribution of doped holes between copper and oxygen, features that the one-band model captures only through effective parameters \cite{White2015threeband}. A wider window means fewer assumptions but a larger local Hilbert space, more entanglement, and a more delicate $d$--$p$ level splitting. Since two groups starting from the same functional and structure can arrive at models with different orbital content, the window and the Wannierization are part of the model and should be reported as such.

The pressurized bilayer nickelate La$_3$Ni$_2$O$_7$ is a case in which the downfolding is still being worked out. Density functional calculations at the pressure where superconductivity appears give two Ni $e_g$ bands per layer, with the $d_{z^2}$ orbitals split into bonding and antibonding combinations by the shared apical oxygen. From this common starting point a bilayer two-orbital Hubbard model was derived \cite{Luo2023}, constrained random-phase interactions were fed into a DMFT study \cite{Christiansson2023}, and a further reduction that treats the half-filled $d_{z^2}$ orbitals as local moments coupled across the bilayer and the $d_{x^2-y^2}$ orbitals as itinerant produced the $t$--$J$--$J_\perp$ model that DMRG solved to show how interlayer exchange mediates pairing \cite{Qu2024}. 

For suitable quasi-one-dimensional downfolded models, DMRG can tightly converge even multiorbital solutions, allowing model and input uncertainties to be tested with reduced solver ambiguity. The iron-based ladder BaFe$_2$S$_3$, which superconducts under pressure, was downfolded onto a multiorbital Hubbard--Kanamori model on a two-leg ladder, and DMRG established its magnetic order and pairing tendencies \cite{Patel2016}; for the related selenide, the dynamical spin structure factor of the multiorbital model agrees qualitatively with the inelastic neutron spectrum of the block orbital-selective Mott phase \cite{Herbrych2018}. Orbital-selective localization requires a multiorbital description, which also captures the coupling of localized and itinerant electrons implicated in the block magnetic order. The layered organic salts $\kappa$-(BEDT-TTF)$_2X$ are downfolded onto a triangular-lattice Hubbard model whose $t'/t$ and $U/t$ span the range in which a putative spin-liquid regime between the metal and the ordered Mott insulator has been studied with cylinder DMRG and finite-temperature tensor networks \cite{Nakamura2009,Szasz2020,Wietek2021}. A change of a few percent in $t'/t$, within the spread between structures, changes the predicted phase. 

The practical response is to treat the downfolded Hamiltonian as an ensemble rather than a point. Constrained random-phase interactions can vary substantially with the choice of orbital window \cite{Vaugier2012}, while the La and Hg cuprates illustrate structural dependence of the effective parameters \cite{Hirayama2018}. Different windows define different effective models, and differences between compounds are physical variation, not error bars on one Hamiltonian. Both motivate testing the robustness of a phase assignment to defensible model and input choices \cite{Kent2018}. Solving a small set of defensible Hamiltonians is affordable when each solution is cheap, and when it is not, reduced density matrices can give local energy sensitivities to hoppings and interactions (Sec.~\ref{sec:uncertainty}). A pairing pattern or phase boundary that survives the ensemble is a stronger material conclusion than one obtained by converging the bond dimension for a single parameter set. Downfolding into few-band models of quasi-one-dimensional and layered compounds is an \emph{advanced specialist} capability; fully multiorbital two-dimensional models with material-specific interactions are \emph{emerging}; and an automated loop that re-derives the downfolded model in response to disagreement with experiment is not yet routine.

\subsection{Dynamical mean-field theory with tensor network impurity solvers}\label{sec:dmft}

Dynamical mean-field theory (DMFT) maps a lattice problem onto a self-consistent impurity coupled to a bath and, combined with density functional theory, is the standard route to the spectra of correlated materials \cite{Georges1996,Kotliar2006}. The impurity is the correlated shell of one atom, three to seven orbitals with their full Coulomb, Hund's, and spin--orbit terms; the bath is a hybridization function $\Delta(\omega)$ that summarizes the rest of the crystal and is updated until the impurity and local lattice Green functions agree. The bottleneck is the impurity solver. Continuous-time quantum Monte Carlo, the workhorse solver, returns imaginary-time data whose continuation to real frequencies is ill-posed, can develop a sign problem with spin--orbit coupling, off-diagonal hybridization, or general interactions, depending on the model and basis, and slows down as the temperature is lowered. Tensor networks enter by discretizing the bath into a finite number of sites per orbital, so that impurity plus bath becomes a finite Hamiltonian on a chain, a star, or a tree. The discretization is controlled rather than ad hoc: a bath of $N$ sites reproduces the continuous hybridization exactly up to a propagation time that grows with $N$, or equivalently down to a frequency resolution set by it \cite{deVega2015}, and the chain form follows from the continuum by a Lanczos transformation \cite{Busser2013}, the same construction as the chain mapping of Sec.~\ref{sec:thermal} \cite{Chin2010chain}. The solver can then work directly on the real-frequency axis at zero temperature, or in imaginary time, and handles general interactions without the conventional sign problem; in exchange the finite bath limits the frequency resolution, most visibly at low frequency, and its cost is set by the entanglement between impurity and bath.

The user faces several design choices with no single right answer. The bath can be a chain, with the impurity at one end and the bath states tridiagonalized, or a star, in which the impurity couples directly to bath modes ordered by energy; the star separates occupied from empty modes, carries less entanglement, and has been favoured in recent real- and imaginary-time solvers \cite{Wolf2014star,Wolf2014,Wolf2015}. The bath can be fitted on the imaginary axis, where the hybridization is smooth and a few dozen sites suffice \cite{Wolf2015}, or on a dense real-frequency grid, which needs more sites but gives spectra whose resolution depends on bath discretization and the spectral algorithm, without requiring analytic continuation. The spectral function is then obtained by a Chebyshev expansion \cite{Ganahl2014,Wolf2014}, by real-time evolution followed by linear prediction \cite{Ganahl2015}, or by imaginary-time evolution when only Matsubara data are needed \cite{Wolf2015}.

For a multiorbital impurity, the fork tensor product state gives each orbital its own bath chain attached to a common impurity, reducing the need to carry correlations between different bath branches along a single chain and making real-time evolution of three-orbital models tractable \cite{Bauernfeind2017}; reducing the bath to a few natural impurity orbitals and solving the resulting small problem with DMRG is another route to a real-frequency solver \cite{Lu2019NO}. The numerical renormalization group is itself a matrix product state method on a logarithmically discretized bath \cite{Bulla2008}: it has unmatched resolution at low frequency but limited resolution at high frequency and a cost that grows exponentially with the number of orbitals, so it and the DMRG-based solvers are complementary.

These choices have been tested on real materials. The fork solver reproduced the quasiparticle peak and Hubbard bands of the three-orbital $t_{2g}$ shell of SrVO$_3$ on the real axis without continuation \cite{Bauernfeind2017}, and applied to a $d$--$p$ model of SrMnO$_3$ it resolved the manganese--oxygen hybridization and the multiplet structure of the Hubbard bands, features that continuation from imaginary time blurs \cite{Bauernfeind2018}. For Sr$_2$RuO$_4$, whose three-orbital impurity with spin--orbit coupling has an off-diagonal hybridization that would produce a sign problem in Monte Carlo, an imaginary-time matrix product state solver gave orbital- and spin-resolved self-energies at low temperature \cite{Linden2020}, and a numerical renormalization group treatment traced how its Fermi liquid emerges from Hund's-metal physics \cite{Kugler2020}. Nonequilibrium DMFT has been solved with matrix product states on a star \cite{Wolf2014star}, and cluster extensions, needed for the momentum-dependent self-energies of the cuprates, include two-site calculations \cite{Wolf2015} and four-site cellular DMFT with a DMRG solver, including finite doping \cite{NunezFernandez2018dmft}. Larger clusters with realistic multiorbital interactions and well-resolved low-energy spectra remain demanding.

Quantics tensor trains add a second, different use of tensor networks inside the same loop. Green functions, self-energies, and hybridization functions on dense imaginary-time or frequency grids are smooth and hence compressible once their coordinate is encoded in binary digits, and the compression persists across the range of scales, from bandwidth to temperature, that a low-temperature calculation must resolve \cite{Shinaoka2023}. Tensor cross interpolation builds the representation from few function evaluations \cite{Ritter2024}, two-particle vertices and the Bethe--Salpeter equations that connect them to susceptibilities fit the same format \cite{Rohshap2025}, and a strong-coupling impurity solver has been formulated directly in it \cite{Kim2025}. The tensor network here compresses correlation functions rather than the wavefunction, and its promise is to make affordable the diagrammatic extensions of DMFT that add nonlocal correlations and have been limited by the size of the vertex.

Beyond the orbital, interaction, and double-counting choices shared with downfolding, the error budget is dominated by the bath discretization and its fit, the DMFT self-consistency, and the frequency resolution of the spectrum. Real-frequency observables should be converged in the number of bath sites and the broadening, with most scrutiny at low frequency, where narrow spectral features are particularly sensitive to bath resolution. Single-site multiorbital DMFT for real materials with matrix product state solvers, including spin--orbit coupling and real-frequency spectra, is an \emph{advanced specialist} capability; cluster DMFT with tensor network solvers, charge self-consistency with the density functional, and the compression of vertex functions for nonlocal extensions are \emph{emerging}.

\subsection{Quantum embedding}\label{sec:embedding}

Quantum embedding keeps an explicit crystalline Hamiltonian but treats only a fragment or an active space at high many-body accuracy \cite{Sun2016embedding}. Density matrix embedding theory (DMET) is the variant that pairs most naturally with a tensor network solver \cite{Knizia2012,Zheng2016}. Where DMFT matches the local Green function through a frequency-dependent bath, DMET matches the one-particle density matrix through a bath that is finite and algebraic: the Schmidt decomposition of a mean-field state across the fragment boundary yields at most as many entangled bath orbitals as fragment orbitals, so a fragment of $L$ orbitals becomes an impurity problem of at most $2L$ orbitals with a Hamiltonian projected from the full crystal. That problem has the size and structure of a molecular active space, and DMRG is the natural solver once $L$ exceeds a dozen orbitals. A correlation potential on the mean field is adjusted to minimize the mismatch between selected mean-field and correlated density-matrix elements, and the energy is assembled from fragment contributions. Standard ground-state DMET targets energies, reduced density matrices, local moments, order parameters, and their trends, and the approximation is improved systematically by enlarging the fragment \cite{Cui2020}; spectral functions require additional response constructions at a cost set by the fragment size.

The ab initio workflow starts from a periodic Hartree--Fock or density functional calculation in a Gaussian basis, takes an atom, a cluster of atoms, or a unit cell with all its orbitals as the fragment, builds the bath from the Schmidt decomposition of the periodic mean-field state, and solves the resulting problem of typically fifty to a hundred orbitals with DMRG \cite{Cui2020}. Applied to the cuprates, it computed the local moments, gaps, and magnetic correlations of the parent compounds and their dependence on the crystal structure \cite{Cui2022}, and followed the trends in pairing across doped compounds \cite{Cui2025}, starting from crystal structures and a Gaussian basis rather than from a fitted model. A different embedding, in which small clusters cut from the crystal are treated with wavefunction methods in a quantum-chemical environment, showed that the superexchange in corner-sharing cuprates is enhanced by the coupling between the oxygen breathing motion and the hopping \cite{Bogdanov2022}, a mechanism that neither a downfolded model with fixed parameters nor a mean-field environment would produce.

The embedding route is also how tensor networks reach the chemistry of extended systems: the metalloenzyme active sites of Sec.~\ref{sec:molecules} can be treated as DMRG active spaces embedded in a classical or low-level quantum description of the protein, and the same construction applies to a transition-metal centre on an oxide surface or in a zeolite pore. In the hydrogen-chain benchmark that compared essentially every many-body method on the same ab initio Hamiltonian, DMRG supplied near-exact reference energies in the minimal basis and, together with multireference configuration interaction, anchored the larger-basis comparisons \cite{Motta2017}. This is the role tensor networks are best placed to play: certifying a workflow they do not run end to end. Embedding of molecular active spaces is \emph{advanced specialist}; ab initio embedding of correlated solids with a DMRG solver is \emph{emerging}, and it is the route with the shortest distance between a crystal structure and a many-body number.

\subsection{Sources of uncertainty from material to observable}\label{sec:uncertainty}

Each of the four routes introduces approximations before the many-body problem is solved. Electronic-structure inputs depend on the basis and, in DFT, on the exchange--correlation functional. Downfolding retains selected orbitals and interactions, often replacing frequency-dependent screening by static couplings. Experimentally constrained models may omit phonons or weak interlayer interactions. Single-site DMFT assumes a local self-energy, while embedding approximates the environment of a finite correlated fragment. Molecular active spaces leave correlations involving excluded orbitals to an additional approximate treatment. Together with uncertainties in structure and sample conditions, these choices can change the predicted phase or excitation energies.

Tensor networks address the many-body solution of the chosen Hamiltonian. They treat its interactions nonperturbatively, and their error is controlled and systematically improvable, since increasing the bond dimension approaches the exact solution. Reliable results also require the checks of truncation, environment, size and geometry, ordering, time and frequency resolution, and optimization collected in Box~1. Such convergence controls the solver's own approximations, not those introduced by model selection and construction. The physical assumptions connecting the chosen Hamiltonian to the material must therefore be validated separately.

Validation should assess how predictions change under plausible variations in material inputs, model parameters, and the degrees of freedom or interactions retained. Local sensitivities can guide which parameters to vary, while differentiable solvers can support sensitivity analysis and fitting to experimental data \cite{Liao2019,Yu2021}. Comparisons with measured spectra should match the experimental temperature and account for probe matrix elements, instrumental resolution, and uncertainties from analytic continuation where required. Reporting which conclusions survive these checks, alongside solver convergence, helps distinguish limitations of the material description from numerical errors.

\section{Where We Stand and Where We Are Going}\label{sec:frontiers}

\subsection{A map of maturity}\label{sec:maturity}

Table~\ref{tab:master} condenses Secs.~\ref{sec:regimes} and \ref{sec:real-materials} into a method-selection guide. For each class of problem it names the typical Hamiltonian, the natural network and algorithm, the observables that come out, the maturity of the capability, and a representative system. The maturity labels apply to a task in a stated setting, not to a network family. Established does not mean cheap, and emerging records demonstrated calculations rather than routine predictive use.

\begin{table}[p]
\centering
\setstretch{1.0}
\scriptsize
\setlength{\tabcolsep}{3pt}
\renewcommand{\arraystretch}{1.15}
\begin{tabularx}{\textwidth}{@{}>{\RaggedRight\arraybackslash}p{0.11\textwidth}>{\RaggedRight\arraybackslash}p{0.14\textwidth}>{\RaggedRight\arraybackslash}p{0.18\textwidth}>{\RaggedRight\arraybackslash}p{0.155\textwidth}>{\RaggedRight\arraybackslash}p{0.145\textwidth}>{\RaggedRight\arraybackslash}X@{}}
\toprule
\textbf{Problem} & \textbf{Typical Hamiltonian} & \textbf{Network and algorithm} & \textbf{Key observables} & \textbf{Maturity} & \textbf{Representative systems} \\
\midrule
Spin chains and ladders & Heisenberg, XXZ, spin--Peierls & MPS: finite and infinite DMRG; TEBD, TDVP & energies, magnetization, gaps, correlations, entanglement spectra, $S(k,\omega)$, thermodynamics & Established & KCuF$_3$; Sr$_{14}$Cu$_{24}$O$_{41}$ \\ \hline
Frustrated magnets in 2D and 3D & Heisenberg on kagome, triangular, $J_1$--$J_2$ square, and pyrochlore lattices; anisotropic exchange & cylinder DMRG; infinite PEPS with CTMRG; isometric networks; SU(2) DMRG on 3D clusters & energies, magnetization and plateaus, correlations, entanglement diagnostics, $S(k,\omega)$ & Advanced specialist (2D); emerging (3D) & kagome 1/9 plateau; Ca$_{10}$Cr$_7$O$_{28}$; KYbSe$_2$; Ba$_3$CoSb$_2$O$_9$; pyrochlore \\ \hline
Doped fermions and superconductivity & Hubbard, $t$--$J$, bilayer and multiorbital variants & cylinder DMRG; fermionic infinite and finite PEPS & stripes, pairing correlations, competing orders & Advanced specialist; controlled 2D $T_c$ open & 2D Hubbard benchmark; La$_3$Ni$_2$O$_7$ \\ \hline
Moir\'e materials & continuum models compressed to MPO, with sample-dependent parameters & cylinder DMRG & correlated insulators, Fermi liquids, fractional Chern states & Advanced specialist & twisted bilayer graphene; twisted MoTe$_2$ \\ \hline
Quasi-1D multiorbital and electron--phonon models & DFT-derived multiorbital Hubbard--Kanamori; three-band Emery; Hubbard--Holstein & DMRG on chains and ladders & orbital-selective phases, $S(k,\omega)$, $A(k,\omega)$, pairing & Advanced specialist & BaFe$_2$(S,Se)$_3$ ladders; 1D cuprate chains; three-band CuO$_2$ ladders \\ \hline
Spectroscopy and dynamics & any of the above plus a perturbation or drive & TDVP, TEBD, Chebyshev, dynamical DMRG, excitation ansatz & $S(k,\omega)$, $A(k,\omega)$, optical response, quench dynamics & Established 1D; specialist ladders, cylinders, impurities; emerging 2D; open long-time 2D & neutron scattering on chain and triangular compounds \\ \hline
Finite temperature and open systems & Hamiltonian plus bath or Liouvillian & purification, METTS, MPDO and PEPO, XTRG, tanTRG, TRG and TNR, TEMPO & thermodynamics, thermal spectra, steady states, transport & Established 1D equilibrium; specialist open systems; emerging 2D; open material-specific devices & 2D Hubbard thermodynamics; $\alpha$-RuCl$_3$; impurity transport \\ \hline
Molecular active spaces & ab initio active-space Hamiltonian; Heisenberg and Hubbard models on molecular graphs & DMRG and DMRG-SCF over ordered orbitals; TTNS, T3NS; SU(2) DMRG & multireference energies, reduced densities, spin-state gaps & Established; trees specialist & Cr$_2$; Fe--S clusters; FeMoco; Mn$_4$CaO$_5$; [Cu$_2$O$_2$]$^{2+}$; Mn$_{84}$ \\ \hline
Vibrational and vibronic structure & mode Hamiltonians; electron--phonon models & mode-adapted MPS and TTNS; local basis optimization & anharmonic spectra, vibronic dynamics & Advanced specialist & acetonitrile; ab initio vibronic models \\ \hline
Impurities in DMFT & Anderson impurity with discretized bath & MPS and fork tensor product states; Chebyshev, imaginary time; quantics tensor trains & $A(\omega)$, self-energies, multiplet weights & Advanced specialist; vertex compression emerging & SrVO$_3$; SrMnO$_3$; Sr$_2$RuO$_4$ \\ \hline
Embedded fragments & ab initio periodic Hamiltonian with fragment and bath & DMET with DMRG solver & local moments, gaps, pairing trends & Emerging (solids); specialist (molecular active spaces) & cuprate family \\
\bottomrule
\end{tabularx}
\caption{Method-selection guide for tensor network materials simulation. Maturity is assigned to the stated task and setting, and the representative systems are those discussed in the text.}
\label{tab:master}
\end{table}
\FloatBarrier

\subsection{Frontiers}\label{sec:open}

Seven developments would move the largest number of entries in Table~\ref{tab:master} up one level. For each we state the obstacle and what would count as success.

\textit{Two-dimensional contraction and optimization.---} Approximate environment contraction is a central limitation of generic PEPS calculations; cylinder MPS methods instead face severe width and entanglement limits. Automatic differentiation, variational infinite PEPS at larger bond dimensions, finite fermionic PEPS that overtake the largest practical DMRG calculations on eight-leg ladders, accelerator hardware, and belief-propagation contraction have all widened the reach \cite{Liao2019,Naumann2025,Liu2025fpeps,Ganahl2023,Menczer2025,Bayraktar2023,Tindall2023bp}, although belief propagation can become unreliable near criticality \cite{Midha2026}. The obstacle is that the environment error, the truncation error, and the optimization path are entangled with one another, so a PEPS result is harder to certify than a DMRG result even when it is more accurate. Success would be a two-dimensional calculation that reports a convergence sequence in both $\chi$ and the environment dimension with the routine confidence of a DMRG discarded-weight extrapolation, and agreement between independently converged PEPS, cylinder DMRG, and neural quantum state calculations on specified regions of the doped Hubbard phase diagram. This is the single development on which the largest number of materials applications depends.

\textit{Dynamics and spectra in two dimensions.---} Entanglement growth limits the accessible time window and hence the frequency resolution, while in generic PEPS evolution repeated environment contraction adds a substantial computational burden; the constructions of Sec.~\ref{sec:spectroscopy} each trade accuracy for longer times or for the thermodynamic limit \cite{Lin2022,Vanderstraeten2015exc,Tindall2026,AriasEspinoza2024}. The target is a momentum-resolved dynamical structure factor of a two-dimensional frustrated magnet at the temperature and energy resolution of a modern neutron spectrometer, so that the comparison of Sec.~\ref{sec:models} can be made without a cylinder extrapolation, and a one-particle spectral function of a doped two-dimensional model at finite temperature at the resolution of photoemission. Achieving these targets with controlled two-dimensional extrapolations is not yet routine; the first would help resolve several questions about frustrated compounds left open in Sec.~\ref{sec:frustrated}.

\textit{Low temperature in two dimensions and fermions in three.---} Thermal tensor networks capture signatures of pseudogap physics in the doped Hubbard model \cite{Li2023tantrg} and the development of pairing correlations in the $t$--$t'$--$J$ model \cite{Sinha2026}, but controlled superconducting transition temperatures in the two-dimensional thermodynamic limit remain a major challenge for the repulsive cuprate models discussed here. Three-dimensional networks exist for thermal bosons, for the dynamics of disordered spins, and, with SU(2) DMRG on finite clusters, for the ground state of the pyrochlore antiferromagnet (Sec.~\ref{sec:frustrated}) \cite{Jahromi2020thermal,Tindall2026,Hagymasi2021}, but direct tensor network treatment of realistic three-dimensional correlated fermions remains limited, restricting applications to bulk oxides, intermetallics, and $f$-electron compounds. Lower temperatures and larger transverse sections demand greater representational capacity, while a three-dimensional spatial cut intersects a surface of bonds. The efficient one-dimensional constructions of Sec.~\ref{sec:thermal} do not transfer straightforwardly to these regimes. Success would be a finite-temperature phase boundary of a doped two-dimensional model with an error bar, and a three-dimensional fermionic calculation that goes beyond what DMFT and its cluster extensions deliver. Until then, three-dimensional materials reach tensor networks through the impurity and embedding routes, which is why those routes matter disproportionately.

\textit{Larger active spaces and dynamic correlation.---} Chemistry needs both a larger static-correlation space and a seamless treatment of the dynamic correlation outside it. Trees, spin adaptation, and perturbative or coupled-cluster corrections each address one side \cite{Baiardi2020,Veis2016tcc}, and the fastest progress is in the automatic selection of active space and orbital ordering from entanglement measures, which removes the choice that most often decides whether a DMRG calculation succeeds. Success would be a workflow that, given a molecule and a target accuracy, chooses the active space, the network topology, the ordering, and the correction without expert intervention and returns an error estimate, and that reaches the metalloenzyme cofactors and transition-metal clusters of Sec.~\ref{sec:molecules} with their full first coordination sphere. For materials this matters because the same workflow is the solver inside the embedding route.

\textit{Uncertainty quantification and the inverse problem.---} The sensitivities of Sec.~\ref{sec:uncertainty} are cheap in principle, but derivatives through an approximate contraction or a truncated time evolution are themselves approximate, responses can be singular or discontinuous at phase transitions, and an ensemble of Hamiltonians multiplies a cost that is already at the limit in two dimensions. Success would be a material prediction reported with an error budget that assigns its uncertainty to input, model interface, solver, and observable separately, and a Hamiltonian for a real compound inferred from its full experimental data set with a tensor network forward model and posterior uncertainties on the couplings. Such comprehensive error budgets are not yet standard. Parameter fitting has been demonstrated with classical forward models \cite{Samarakoon2020} and with tensor networks for thermodynamic data of low-dimensional magnets \cite{Yu2021}; extending these approaches to full spectra with calibrated posterior uncertainties remains a further goal.

\textit{Classes of materials not yet reached.---} The applications in this Perspective concentrate in low-dimensional magnets, layered superconductors, molecules, and single-site impurities, because those are the problems whose entanglement structure a tensor network follows naturally. Disorder and real-space inhomogeneity, ubiquitous in real samples and decisive for many-body localization, glassy magnets, and the cuprates, suit finite tensor networks, which need no translational invariance; two-dimensional many-body localization studies \cite{Kshetrimayum2020mbl} and disordered-spin dynamics \cite{Tindall2026} show that the tools can be adapted to disorder, but their application to specific compounds has barely begun. Interfaces, heterostructures, and surfaces, where the correlated region is a few layers thick and the natural geometry is a slab, offer opportunities for cylinder and PEPS treatments with material-specific Hamiltonians, and catalytic sites on oxide surfaces and in porous frameworks are within reach of the embedding route of Sec.~\ref{sec:embedding}. Progress will require both algorithmic adaptation and collaboration between the tensor network and materials communities. Success would be a first material-specific calculation in each of these classes, reported with the convergence record of Box~1.

\textit{Integration, software, and reproducibility.---} The routes of Sec.~\ref{sec:real-materials} are still assembled by hand, and the assembly is where most of the time and most of the irreproducibility lie. Electronic-structure codes should expose localized orbitals, interactions, windows, and double-counting choices in machine-readable form; tensor network codes should return observables with convergence traces and reduced density matrices; and experimental partners should specify sample conditions and resolution functions. The DMFT frameworks that carry a band structure through the self-consistency loop with interchangeable impurity solvers \cite{Parcollet2015triqs,Aichhorn2016} show what such integration looks like, and the tensor network solvers of Sec.~\ref{sec:dmft} should be connected to them as first-class options. Mature libraries exist across physics (ITensor, TeNPy, quimb, YASTN), chemistry (Block2, CheMPS2, QCMaquis), and accelerators \cite{Fishman2022,Hauschild2018,Hauschild2024,Gray2018,Rams2025,Zhai2023block2,Wouters2014,Szenes2025,Menczer2025}, but their conventions are fragmented \cite{Sehlstedt2026}, and a versioned interchange format for tensor network states would let one code optimize a state and another evaluate it. A community reference workflow that carries one material from structure to measured spectrum, with validation gates at every interface and every intermediate object archived, would be worth more than any single record calculation; a chain or ladder compound for which the solver is essentially exact is the natural candidate, so that the workflow rather than the solver is tested.

\subsection{Tensor networks, quantum computers, and machine learning}\label{sec:qc-ai}

Two neighbouring fields---quantum computing and machine learning---have also shaped how tensor networks are used today. 

Tensor networks are related to quantum computing in two ways. First, a quantum circuit is itself a tensor network, so the action of a quantum circuit on an initial state maps to a tensor network contraction. Thus, tensor network contraction algorithms lend themselves to simulating quantum circuits naturally. Second, certain tensor network states can be readily compiled into quantum circuits; these tensor networks can easily integrate into and potentially enhance some quantum algorithms. On the circuit simulation side, cost depends on circuit connectivity, depth, entanglement, and the requested output as well as qubit count \cite{Markov2008circuit,Berezutskii2025qc}. Tensor networks therefore provide important tests of proposed quantum computational advantages, although their failure alone does not establish classical intractability. Belief-propagation and sparse tensor network methods reproduced and then surpassed the kicked-Ising dynamics of a 127-qubit processor \cite{Tindall2024,Begusic2024}, graph-based PEPS that follow the native connectivity of the hardware reached the largest processors available \cite{Patra2024}, and two- and three-dimensional tensor network dynamics has re-examined the more recent annealing claim of Ref.~\cite{King2025} \cite{Tindall2026}, although some experiments have so far resisted classical reproduction \cite{GoogleQE2025,Bermejo2026}; the state of this contest is reviewed in \cite{Kshetrimayum2026}. It matters for materials because the same devices are proposed as simulators of the Hubbard and Heisenberg models discussed above, and tensor networks provide a demanding classical baseline and can validate results within their accessible regimes.

On the algorithmic side, matrix product states can guide circuit preparation of correlated initial states, with depth and fidelity depending on bond dimension and the circuit construction \cite{Ran2020,Rudolph2023,Rudolph2024}, while sample-based quantum diagonalization uses hardware samples to select determinants for classical diagonalization \cite{RobledoMoreno2025}.

The relation between tensor networks and machine learning runs in both directions. Neural quantum states compete with tensor networks as variational ans\"atze, and hybrids between the two are an active frontier (Sec.~\ref{sec:positioning}) \cite{Lange2024,Chen2024nqs}. In the other direction, tensor decompositions compress the parameter arrays of neural networks and large language models \cite{Novikov2015,Tomut2024,Jahromi2024tnn}, serve as generative models in their own right \cite{Han2018}, and provide equivariant interatomic potentials for atomistic simulation \cite{Hodapp2024}. Whether these yield more than conventional low-rank compression depends on the rank the data actually have, and they remain an outlook rather than an established advantage.

\section{Conclusion}\label{sec:conclusion}

Tensor networks began with the density matrix renormalization group, which grew out of attempts to apply Wilson's numerical renormalization group to spin chains \cite{White1999}, and have become a general language for many-body states, operators, and thermal ensembles whose computational reach depends strongly on entanglement, geometry, system size, and the observable of interest. For the materials-science community, three conclusions follow. First, the methods already deliver mature results for a well-defined set of questions: ground states and spectra of low-dimensional magnets and large active spaces in molecules, and specialist access to two-dimensional phase competition, to finite-temperature and dynamical properties beyond one dimension, and to impurity solvers for dynamical mean-field theory. These successes are conditional on the Hamiltonian, the geometry, the observable, and the convergence evidence, and Table~\ref{tab:master} records those conditions. Second, different problems call for different networks. A chain, a branched active space, a two-dimensional lattice, a thermal operator, and an impurity bath should not be forced into one representation or judged by one error proxy; the geometry should follow the entanglement, and the convergence protocol should follow the geometry. Third, the decisive frontier is the transition from controlled models to reproducible material-specific prediction. Where solver convergence is demonstrated, the dominant uncertainty in a materials claim can enter upstream, in the structure, the downfolding, or the embedding, or downstream, in the conversion of a correlation function into a measured spectrum.

Crossing that frontier requires communities that have worked in sequence to work together: electronic-structure researchers who define orbitals and interactions, tensor network developers who supply scaling diagnostics and portable convergence records, quantum chemists and embedding researchers who expose fragment and bath choices, experimentalists who specify the conditions of comparison, and software teams who connect all of these through interoperable formats and reference workflows. The measure of progress is a calculation that another group can reproduce, perturb, and potentially falsify at the level of a measured quantity. That standard will not make every correlated material tractable, but it can turn tensor networks from powerful specialist solvers into dependable components of predictive materials science.

\section*{Acknowledgments}
The concept for this Perspective emerged from discussions at the AI4AM 2026 conference, which helped shape its focus for the materials-science community. We thank Multiverse Computing and the Donostia International Physics Center for institutional support. 

\section*{Author contributions}
All authors contributed to the conception, writing, and revision of the manuscript.

\section*{Data availability statement}
No new data were created or analysed in this study.

\section*{ORCID iDs}
Saeed S.~Jahromi: \href{https://orcid.org/0000-0001-7994-8192}{0000-0001-7994-8192}\\
Augustine Kshetrimayum: \href{https://orcid.org/0000-0002-0130-1142}{0000-0002-0130-1142}\\
Roman Rausch: \href{https://orcid.org/0000-0002-0427-3634}{0000-0002-0427-3634}\\
Sukhbinder Singh: \href{https://orcid.org/0000-0002-3099-4179}{0000-0002-3099-4179}\\
Rom\'an Or\'us: \href{https://orcid.org/0000-0002-4496-8115}{0000-0002-4496-8115}

\section*{Competing interests}
The authors declare no competing interests.

\begingroup
\sloppy
\hbadness=2000
\singlespacing
\small
\bibliographystyle{unsrtnat}
\bibliography{references}
\endgroup

\end{document}